# Describing Colors, Textures and Shapes for Content Based Image Retrieval – A Survey

Jamil Ahmad[1], Muhammad Sajjad[1], Irfan Mehmood[1], Seungmin Rho[2], Sung Wook Baik[1*]

Tel: +82-02-3408-3797, Fax: +82-02-3408-4339

jamil.ahmad@icp.edu.pk, sajjad@sju.ac.kr, irfanmehmood@sju.ac.kr, smrho@sungkyul.ac.kr, sbaik@sejong.ac.kr

[1]*College of Electronics and Information Engineering, Sejong University, Seoul, Korea*
[2]*Department of Multimedia, Sungkyul University, Anyang, Korea*

*Abstract*

*Visual media has always been the most enjoyed way of communication. From the advent of television to the modern day hand held computers, we have witnessed the exponential growth of images around us. Undoubtedly it's a fact that they carry a lot of information in them which needs be utilized in an effective manner. Hence intense need has been felt to efficiently index and store large image collections for effective and on-demand retrieval. For this purpose low-level features extracted from the image contents like color, texture and shape has been used. Content based image retrieval systems employing these features has proven very successful. Image retrieval has promising applications in numerous fields and hence has motivated researchers all over the world. New and improved ways to represent visual content are being developed each day. Tremendous amount of research has been carried out in the last decade. In this paper we will present a detailed overview of some of the powerful color, texture and shape descriptors for content based image retrieval. A comparative analysis will also be carried out for providing an insight into outstanding challenges in this field.*

## 1. Introduction

The dawn of the twenty first century brought with itself limitless opportunities and challenges. It is rightly said that $21^{st}$ century is the age of information. With the prevalent trend of capturing almost everything including text, images, music and videos in digital form in this modern age of digital computers, the huge data collections are becoming hard to handle [1]. Thus giving rise to challenges regarding their efficient storage, indexing and on-demand retrieval. The presence of complex patterns in visual data is a major hurdle in developing efficient retrieval systems for multimedia data collections. During the last decade, this challenge has been met with tremendous endeavors by researchers all around the globe. This work focuses on the notable accomplishments in the field of image indexing and retrieval systems during the last decade.

Image retrieval is the process of retrieving similar looking images from large collections usually by submitting a sample image as a query [2]. The retrieval system searches for images that are visually similar to the query image and returns a list of images ranked in order of similarity to the query image[3]. Generally there are two ways for extracting relevant images from large image collections. The traditional annotations based image retrieval (ABIR) and the recent one termed as Content based image retrieval (CBIR). ABIR methods generally rely on manual association of tags to images when they are stored in some database. Then the traditional text based retrieval systems are employed to retrieve images by searching keywords specified in the text query in the tags associated with images in the dataset [4]. The exponential growth of the images around us is making it extremely difficult to manually associate tags with each and every image. Secondly it is very difficult to describe visual content in words. Sometimes an image can simply not be described in words [5]. Thirdly there exist no general rule for describing visual content in textual form. So the description by one person might never be understood by others thus making it extremely difficult for users to get effective output from such text-based image retrieval systems [1].

---

* Corresponding author: E-mail: sbaik@sejong,ac.kr

To avoid the need of manual annotations and to overcome some of the issues in ABIR, content based image retrieval methods were introduced in the early 90s [2]. In CBIR, low-level features extracted from image contents are used to index and search images. A feature is any characteristic of the image that can be measured. Thus features are a collection of numbers derived from image color, texture and shape of objects contained in the image as shown in figure 1. Numerous methods exist in the literature for describing these features and even newer methods are being introduced regularly by the researcher community from the academia as well as the industry. These low-level features has been extensively used for CBIR during the last two decades. However, these low-level features fail to describe the semantics of the image. This difference in associating low-level features with high level semantics is a very well-known problem termed as the "semantic gap" [5]. Efforts are in progress targeting the reduction of this so called semantic gap. Machine learning methods are generally applied for deriving high level semantics from the low-level features. However the limitations in current artificial intelligence and other related techniques are a major hurdle in the way of the researchers. Hence the gap still exists and image retrieval systems are plagued with it affecting their performance badly.

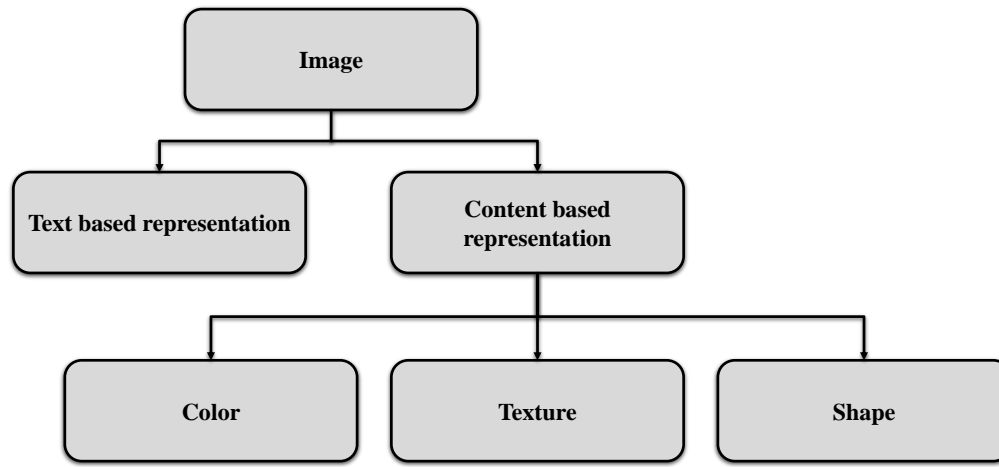

**Figure 1.** describing image content

A typical CBIR system consist of a number of modules including query specification interface, features extraction modules, features representation and indexing modules, searching modules and image retrieval and ranking modules. In addition to these, there might exist machine learning techniques integrated into the feature extraction module [6] for deriving high level semantic description from low-level features and a relevance feedback mechanism as illustrated in figure 2. The purpose of all of these modules, the challenges at all of these stages of image retrieval are being discussed in the subsequent sections.

## 2. Low-Level Image Features

A digital image is merely a collection of pixels represented as large matrices of integers corresponding to the intensities or colors of pixels at different positions in the image [7]. The general purpose of feature extraction is to extract meaningful patterns from these numbers. These patterns can be found in colors, textures and shapes. The measurements derived from these patterns are referred to as low-level features [8]. Methods developed recently for extraction of such features are highlighted in the subsequent sections.

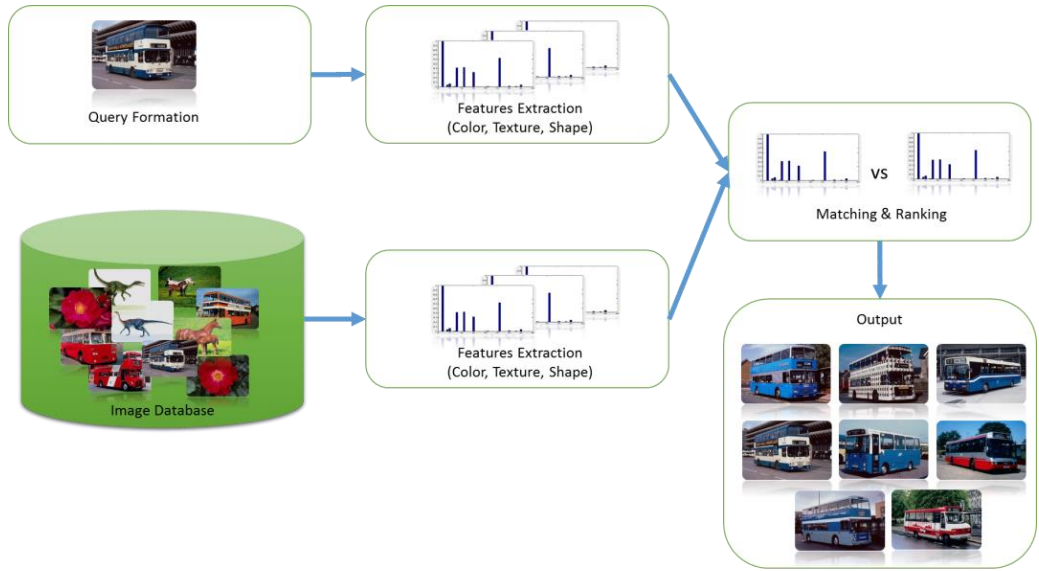

**Figure 2.** Typical CBIR systems

**2.1 Color Features**

The human visual system is very sensitive to colors and possesses the capability to differentiate between thousands of colors. Color sensation is a prime feature of our visual perception system that guides us towards object recognition. Hence it has been widely exploited for representing image content. Colors in digital images may be represented in a variety of color models. These color models are 3-dimensional coordinate systems for representing them as points in that space. Numerous models exist including RGB, HSV, YCbCr and CIELab etc [9]. All of these color representations have been utilized for color based content description. The simplest way to represent colors in an image is to populate color histograms in which a count of the number of pixels of various colors is accumulated [10-12]. Color quantization is generally employed to reduce the number of colors in the image into a few representative colors. This helps making the description process convenient and efficient by reducing memory requirements. The problem with such histograms is that they carry absolutely no information regarding the spatial arrangement of those colors in the image. Hence methods were introduced in an attempt to incorporate some spatial layout information. Color correlograms (CCG) [13] encode color layout as probabilities of occurrence of pairs of colors at a fixed distance in the image. Hence CCG performs better than color histograms. Several color descriptors have also been introduced in MPEG-7 standard. In Dominant Color Descriptor (DCD) [14], colors in the image are quantized into a number of dominant colors using clustering methods. It then encodes dominant colors into triads of color values, percentages of colors and their variances. The drawback associated with this scheme is the inability of represent spatial relationship of colors. The Color Structure Descriptor (CSD) [14] populates a histogram by taking local distributions of colors in the image. A small window is moved from pixel to pixel in the entire image. The histogram bins associated with colors located in the small window are incremented. In this way spatial layout of colors in the image is captured. The Scalable Color Descriptor (SCD) [14] quantizes the color image into 256 colors in the HSV color space. The hue component (H) is quantized into 16 levels. Whereas the remaining two components saturation (S) and value (V) are quantized into 4 components each. A novel Haar transform encoding is then used which facilitates scalable representation of the description. Also provides complexity scalability to the feature extraction and matching procedures. The Color Layout Descriptor (CLD) [14] represents the spatial arrangement of the dominant colors in the region of interest on a regular grid. It is a compact and efficient descriptor for browsing and searching. It is suitable for both still images and videos.

The MPEG-7 color descriptors were further improved by researchers during the last decade. Weaknesses were discovered in their representations and improvements were brought into them. Dominant Color Structure Descriptor (DCSD) [15] combines the characteristics of DCD and CSD into a single descriptor. The image is first quantized into dominant colors. Then the local spatial color distribution is captured by moving a small window over the image and incrementing the corresponding color bins. Unlike DCD, the DCSD represents color structure bins instead of dominant colors in the triads. In another scheme called the Weighted Dominant Color Descriptor (WDCD) [16] adds weights to dominant colors by taking into consideration whether the color contributes to the salient object(s) in the image or the background. These weights represents the importance of a dominant color in the image. A high weight is assigned to the important colors in the image and contributes the most while comparing images. The colors having low weights are assumed to belong to the background and hence contributes the least during the image matching phase. This improved retrieval accuracy of the CBIR system significantly. The WDCD descriptor was tested on natural images as well as cartoon images.

**2.2 Texture Features**

Like colors in the image, the textural characteristics are also effective ways of describing visual content. Texture features have also been widely used in CBIR applications. Textures can be quantified using three principal approaches. These are statistical, structural and spectral [17]. Statistical methods tend to define textures as smooth, coarse, fine, and grainy etc. Simple statistical descriptors include mean, moments, uniformity, correlation, contrast, homogeneity and entropy etc [18]. Structural methods deal with capturing the spatial arrangement of texture primitives in the image. Spectral techniques are derived by analyzing frequency spectrum of the image for global periodicity, high energy and narrow peaks etc [17].

In [19] the author proposed six texture features including directionality, coarseness, contrast, line-likeliness, regularity and roughness. The first three of these features were decided to be more important than the rest after experiments. The QBIC system [20] also uses the histograms of these features. Gabor features have also been used for texture analysis tasks. The responses from Gabor filter banks characterizes the textures.

Texture features introduced in MPEG-7 are Texture Browsing Descriptor (TBD) [21], Homogeneous Texture Descriptor (HTD) [22] and Edge Histogram Descriptor (EHD) [23]. The TBD captures texture characteristics like directionality, coarseness and regularity. It is computed from a set of band-pass filtered images. The filtered output is then used to compute components of the TBD. The descriptor is quite useful for browsing images. In combination with HTD, it can also be employed in retrieval applications with high accuracy. The HTD encodes quantitative characteristics of textures. This descriptor is calculated by applying a bank of filters on the image just like the TBD followed by computing the mean and standard deviation of the outputs in the frequency domain. It has been successfully applied for large scale texture retrieval purposes. The EHD represent textures by local edge characteristics. The spatial distribution of edges is a good texture representation useful for image to image matching even if their texture is not homogeneous. Its computation is fairly simple. The given image is sub-divided into 4 x 4 sub-images. Then local edge histograms for each of these sub-images is computed. Edges are categorized as vertical, horizontal, 45 diagonal, 135 diagonal and non-directional. Each of these type of edge contribute to a bin in the local edge histogram. The entire image gets divided into 16 partitions in which each partition has a five-bin histogram. Thus the full descriptor size is 80 bins.

Most recently local texture features have been exploited for computing global description of the textures. One such method is called the local binary pattern (LBP) [24]. The LBP and its variants have been successfully applied to the task of texture classification. LBP is invariant to illumination changes in the image and since its inception it has been greatly improved. A large number of variants have been derived since then. It extracts information from the textures that is invariant to local grayscale changes in the image. It is calculated at each pixel position taking into account the values of the neighboring pixels forming small circular neighborhoods (with radius R pixels) around the value of a central pixel. The procedure for LBP computation is given in figure 3.

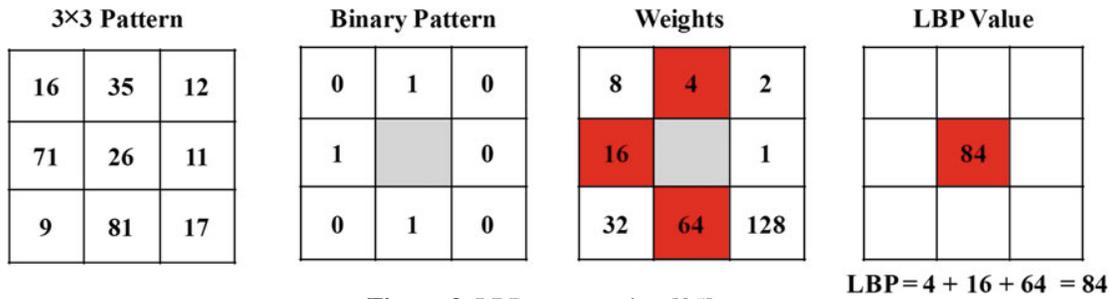

**Figure 3.** LBP computation [25]

The procedure is quite simple. If the value of the surrounding pixel is greater than the pixel value at the center, it produces '1' otherwise outputs '0'. The string of eights bits is converted into a decimal value and put in the position of the central pixel in the LBP map.

A variant of LBP called Centralized Binary Pattern (CBP), proposed in [26], compares pixel pairs in the locality to reduce the lengthy histogram, and it considers the local effect by assigning the largest weight to the central pixel. In another variant Completed LBP (CLBP) [24] utilizes both the magnitude and sign of the difference of the surrounding pixel and the central pixel in contrast to conventional LBP which only utilizes magnitude information. The LBP is a non-directional first order derivative operator. Local Derivative Pattern (LDP) [27] is another variant that uses higher order derivatives. Higher order derivatives capture direction information in the neighboring pixels. The local ternary pattern (LTP) [28] extends the LBP by using three values i.e. -1, 0 and 1 instead of binary values. In local quinary pattern (LQP) [29] five labels are used i.e. -2, -1, 0, 1, 2 based on the amount of difference in the central and surrounding pixel. LBP and its variants have proven to be powerful local descriptors. The extensive adaptation of these methods for a variety of tasks involving texture analysis is a proof of their strength. Hence these methods need to be explored in more detail.

## 2.3 Shape Features

Shape description is an important task in content based image retrieval. It is the process of mapping a 2D shape as a vector such that two similar shapes will have near-to-identical shape descriptors and for dissimilar shapes the descriptors should also be different enough in order to be able to discriminate them. Extensive research is being carried out in the field of shape based image retrieval. This section focuses on the most commonly used shape descriptors derived from the shape contour or shape interior. It also introduces some of the fusion based approaches in this area.

The object's boundary or interior region is commonly used for deriving shape descriptors. Each of the boundary or region based approaches can be further divided into global or structural techniques. Global descriptors use the entire shape to derive shape descriptors whereas the structural approach breaks down the shape into components called primitives and then use those primitives to derive descriptors. A categorization of the existing shape descriptors has been depicted in the figure 4.

The efficiency and effectiveness of shape descriptors is a major challenge in shape based image retrieval. A shape descriptor need to be accurate in retrieving similar shapes from the database. Also the descriptor derivation and matching process should be fast so that it may also support online retrieval of images.

The efficiency and effectiveness of shape descriptors is a major challenge in shape based image retrieval. A shape descriptor need to be accurate in retrieving similar shapes from the database. Also the descriptor derivation and matching process should be fast so that it may also support online retrieval of images.

The following sections summarize both the contour based and boundary based approaches along with their strengths and weaknesses.

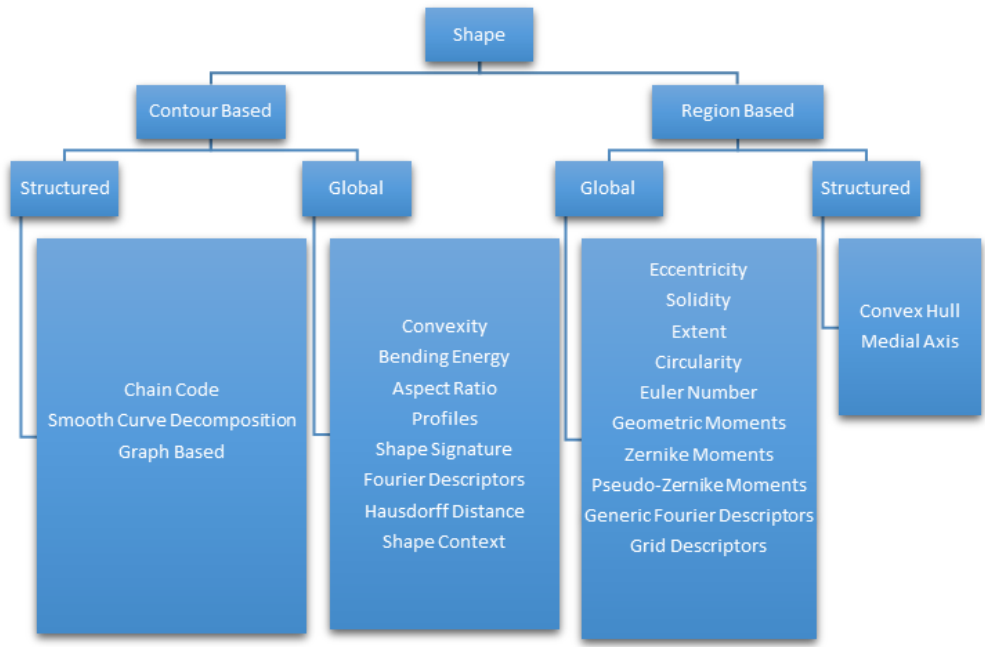

**Figure 4.** Classification of Contour & Region based shape descriptors

**A.  Region based Shape Descriptors**

The following are the shape descriptors that are derived from the entire set of pixels that make up an object. These descriptors can be global or structural.

*Global Shape Descriptors*

Simple geometric characteristics are used to represent shape regions. Such simple descriptors can be used to distinguish shapes having large differences; hence, they are mainly used to exclude false hits or are used in conjunction with other descriptors to describe shapes. These descriptors are basically properties having physical meaning but they are not fit for use as standalone shape descriptors. These include Eccentricity, Solidity, Extent, Circularity and Euler Number.

*Geometric Moments*

Moments are basically the quantitative measure of the shape of a set of points. These moments are also known as Invariant Moments. They are the simplest to compute. Geometric moment function $m_{pq}$ of order (p+q) are given by:

$$m_{pq} = \sum_x \sum_y x^p y^q f(x, y) \quad p, q = 0, 1, 2 \ldots \quad (1)$$

The translation invariant geometric central moments are defined as:

$$\mu_{pq} = \sum_x \sum_y (x - \bar{x})^p (y - y)^q f(x, y) \quad p, q = 0, 1, 2 \ldots \quad (2)$$

Where $\bar{x} = m10/100$ and $\bar{y} = m01/m00$

A set of seven geometric moments are defined by [30]

$$\Phi_1 = \eta_{20} + \eta_{02}$$
$$\Phi_2 = (\eta_{20} - \eta_{02})^2 + 4\eta_{11}^2$$
$$\Phi_3 = (\eta_{30} - 3\eta_{12})^2 + (3\eta_{21} - \eta_{03})^2 \quad (3)$$
$$\Phi_4 = (\eta_{30} + \eta_{12})^2 + (\eta_{21} + \eta_{03})^2$$
$$\Phi_5 = (\eta_{30} - \eta_{12})(\eta_{30} + \eta_{12})[(\eta_{30} + \eta_{12})^2 - 3(\eta_{21} + \eta_{03})^2] + 3(\eta_{21} - \eta_{03})(\eta_{21} + \eta_{03}).[3(\eta_{30} - \eta_{12})^2 - (\eta_{21} + \eta_{03})^2]$$
$$\Phi_6 = (\eta_{20} - \eta_{02})[(\eta_{30} + \eta_{12})^2 - (\eta_{21} + \eta_{03})^2] + 4\eta_{11}^2(\eta_{30} + \eta_{12})(\eta_{21} + \eta_{03})$$
$$\Phi_7 = (3\eta_{21} - \eta_{03})(\eta_{30} + \eta_{12})[(\eta_{30} + \eta_{12})^2 - 3(\eta_{21} + \eta_{03})^2] + (3\eta_{12} - \eta_{03})(\eta_{21} + \eta_{03}).[3(\eta_{30} + \eta_{12})^2 - (\eta_{21} + \eta_{03})^2]$$

Where $\eta_{pq} = \mu_{pq} / \mu_{00}^\gamma$ and $\gamma = 1 + (p+q)/2$ for p+q = 2,3, . . .

Geometric moments are simple to calculate and are invariant to translation, rotation and scaling. But carries several drawbacks as well including information redundancy, sensitivity to noise and great difference in the dynamic range of values.

*Zernike Moments*

Geometric moments were non-orthogonal. Zernike moments (ZM) are orthogonal moments. The complex ZM are derived from the complex Zernike polynomials.

$$V_{nm}(x, y) = V_{nm}(r\cos\theta, \sin\theta) = R_{nm}(r)\exp(jm\theta) \quad (4a)$$

Where $R_{nm}(r)$ is the orthogonal radial polynomial;

$$Rnm(r) = \sum_{s=0}^{(n-|m|)/2} (-1)^s \frac{(n-s)!}{s! \times (\frac{n-2s+|m|}{2})! (\frac{n-2s-|m|}{2})!} . r^{n-2s} \quad (4b)$$

n = 0, 1, 2, …, 0 ≤ |m| ≤ n; and n - |m| is even.

Zernike polynomials are a whole set of complex valued functions orthogonal over the unit disk, i.e., $x^2 + y^2 \leq 1$. The ZM of order *n* with repetition *m* of shape region *f(x, y)* is given by:

$$Znm = \frac{n+1}{n} \sum_r \sum_\theta f(r\cos\theta, r\sin\theta).Rnm(r).\exp(jm\theta) \quad r\leq 1 \quad (5)$$

Zernike moments have several advantages as compared to geometric moments. ZMs have minimal redundancy because of being orthogonal. They are robust to noise. And are invariant to rotation. Besides these advantages ZMs also possess a number of weaknesses like high computational complexity, coordinate space normalization and numerical approximation of the continuous integrals causes errors.

Zernike moments are proven to be very effective in retrieval of similar shapes from large database of images by [31].

*Pseudo-Zernike Moments*

Pseudo-Zernike moments are derived from Pseudo-Zernike Polynomials [32] which are identical to Zernike polynomials but have a different real-value radial polynomial $\tilde{R}_{pq}(r)$ which is defined as

$$R_{pq}(r) = \sum_{k=0}^{p-q} (-1)^k . \frac{(2p+1-k)!}{k!(p-|q|-k)!(p+|q|+1-k)!} . r^{p-k} \quad (6)$$

Where $p \geq 0$, and *q* is a positive or negative integer subject to $|p| \leq q$ only. Pseudo-Zernike moments are calculated as

$$PZM_{pq} = \frac{p+1}{n} \sum_r \sum_\theta f(r\cos\theta, r\sin\theta).R_{pq}(r).e^{-iq\theta} \quad , r\leq 1 \quad (7)$$

A fast and effective algorithm for the computation of these moments is described in [33].

*Generic Fourier Descriptors (GFD)*

These are region based descriptors. These were derived by [34]. The shape was first transformed to polar coordinates and then the 2D Fourier transform was applied for the derivation of the shape descriptor. GFD are calculated as

$$GFD(\omega 1,\omega 2) = \sum_{r=0}^{m-1}\sum_{\theta=0}^{n-1} f(r,\theta)\exp[-j(\frac{\omega_1 r}{m}+\frac{\omega_2 \theta}{n})] \qquad (8)$$

The GFD are invariant to scale, rotation and translation.

*Grid Descriptors (GD)*

Grid descriptors are derived by overlaying a shape on a fixed sized grid. Then the grid cell that is occupied at least 15 % is represented as 1 and the rest are represented as 0s. This yields a matrix of 1s and 0s. This matrix is also called shape matrix. The matrix is then scanned from left to right and in a top-down manner to get a string of 1s and 0s. This binary sequence is the grid descriptor. It is invariant to translation but is not invariant to scale and rotation. So this approach to grid based shape description is modified in different ways [35-37].

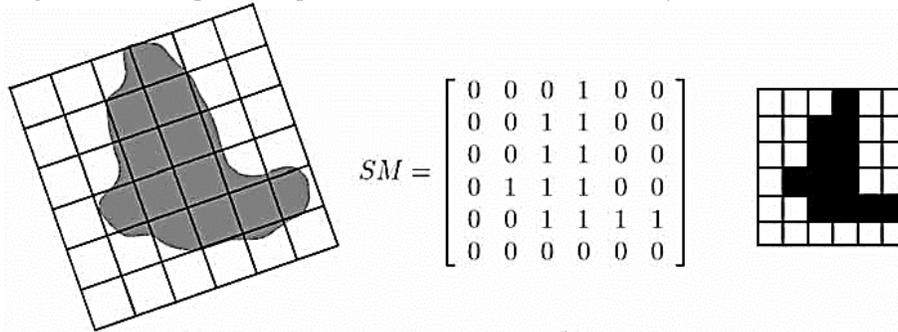

**Figure 5.** Grid based shapes representation

**Structural Shape Descriptors**

This section introduces some of the commonly used region based structural shape descriptors.

*Convex Hull*

In this approach a series of convex hulls is used to represent shape. The convex hull is the smallest convex polygon that entirely contains an object. Before applying this approach, the object boundary is smoothed.

Shape representation is achieved by a recursive procedure which yields a concavity tree as shown in the Figure 6. Each concavity is then described by its area, cord length, max curvature, and distance from max curvature point on the chord. For such representation, shape matching becomes a string matching or graph matching.

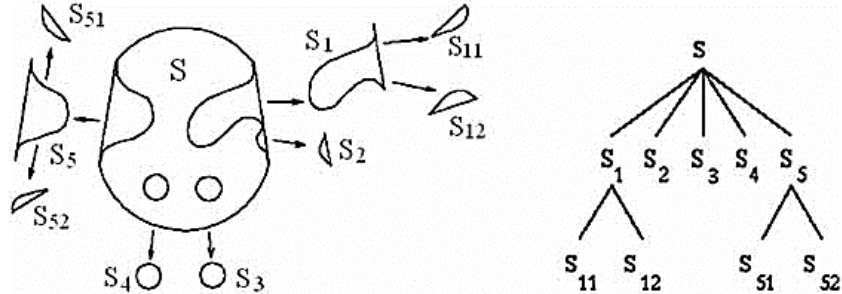

**Figure 6.** Convex Hull based shapes representation and shape tree derived from the convex hull

Its storage efficiency is very high and provides invariance to rotation, scale and translation. It is also robust against noisy boundaries because of the smoothing process involved. However it is quite difficult to extract this convex hull representation of a shape.

*Medial Axis*

A shape can also be represented by a region skeleton. A skeleton is defined as "the connected set of medial lines along the limbs of the figure" [38]. Skeletons and medial axes have been widely and effectively used for shape description using basic lines and arc patterns based structures. It is determined using an image processing procedure which shrinks input shapes to axial stick-like representations of the shape. It is as the loci of centers of bi-tangent circles that fit completely within the foreground region of the shape being considered.

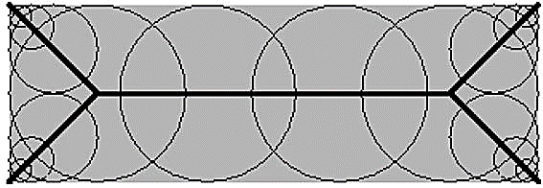

**Figure 7.** Medial Axis representation

The bold line is the shape skeleton. It can be broken down into primitives and represented as graphs according to a certain condition. Hence shape matching becomes graph matching. The calculation of medial axis is also a challenging task. It is also very sensitive to boundary noise. Hence shape contours are smoothed prior to the computation of the medial axis.

### B. Boundary based Shape Retrieval Descriptors

These descriptors are derived by considering only the boundary of the shape. Many techniques have been introduced in the literature for deriving shape descriptors from the shape boundary. This section provides a brief overview of both global and structural boundary descriptors.

Using the geometrical properties of the shape boundary, shapes can be represented as scalar values corresponding to those geometric properties [32]. A few of the commonly used descriptors include Convexity, Bending Energy, Aspect Ratio and Profiles.

*Shape Signatures*

These are one-dimensional functions derived from the shape's contour. A shape signature can alone be used to describe a shape. As well as it can also be used with other descriptors like Fourier Descriptors and Wavelet descriptors. This section introduces some of the common shape signatures.

A. Centroid Distance Function (CDF)

It is defined as the distance between the boundary points and the shape centroid [39]. The shape centroid is basically the central point of a shape and is calculated as

$$Centroid = \begin{cases} gx = \frac{1}{N}\sum_{i=1}^{N} x_i \\ gy = \frac{1}{N}\sum_{i=1}^{N} y_i \end{cases} \quad (9)$$

The distance between the boundary points and the centroid is calculated as

$$r(n) = [x(n) - gx)^2 + (y(n) - gy)^2]^{\frac{1}{2}} \quad (10)$$

The centroid distance function is invariant to translation by definition and it can also be made invariant to scale by normalization.

B. Chord Length Distance

It is also derived from the shape's contour. Unlike the Centroid Distance it does not consider any reference point, instead it calculates the distance between any two boundary point a, b such that distance vector ab is perpendicular to the tangent vector at a [40].

### C. Angular Function

It represents the shape contour as a 1D function of changes in the direction of the shape boundary. These changes are considered to be of importance to human visual system and hence can be used to represent shapes. The angular function at different boundary points is given by

$$\phi(u) = \arctan(\frac{y(u) - y(u-w)}{x(u) - x(u-w)}) \qquad (11)$$

Where w is the step of selected boundary length.

This signature is sensitive to noise. This problem is overcome by filtering the shape before signature computation [32].

### D. Triangular Centroid Area (TCA)

As we proceed along the contour of a shape from point a to point b a triangle can be formed by these two points on the boundary and the shape centroid. The area of all the triangles made by any two successive boundary points and the centroid gives us the shape signature. It can be calculated as [41]

$$TCA(u) = \frac{1}{2} | x_1(u) y_2(u) - x_2(u) y_1(u) | \qquad (12)$$

Where (x1,y1) and (x2,y2) are the two points on the shape boundary.

### E. Triangle Area Representation (TAR)

This shape signature is different from TCA because it forms a triangle by taking all three points from the shape's contour. Hence it is the area formed by these triangles. This area is zero for straight line segments, positive for convex and negative for concave contours [32].

### F. Complex Coordinates

This signature is basically the sequence of complex numbers generated form the contour coordinates as

$$z(n) = [x(n) - gx] + i[(y(n) - gy] \qquad (13)$$

Where (gx, gy) is the shape centroid.

### G. Farthest Point Distance

This shape signature is derived by capturing distances between shape corners. The value of this shape signature at any point a is given by the distance between a and the farthest boundary point b. This shape signature is calculated by adding the Euclidean distance between a and the shape centroid c to the distance between b and c. It is calculated as

$$FPD(u) = \sqrt{([x(u) - xc]^2 + [y(u) - yc]^2} + \sqrt{([x\,fp(u) - xc]^2 + [yfp(u) - yc]^2} \qquad (14)$$

Where $(x_{fp}, y_{fp})$ denotes the farthest point from the boundary point $(x,y)$ and $(xc,yc)$ is the shape centroid.

*Fourier Shape Descriptors*

The Fourier shape descriptors are calculated by applying Fourier Transform on 1D shape signatures discussed previously which are usually derived from the shape's contour. For a shape signature z(u), the Fourier Transform can be calculated as

$$a_n = \frac{1}{N} \sum_{u=0}^{N-1} z(u).e^{-j2\pi nu/N} \quad , n=0,1,...N\text{-}1 \qquad (15)$$

The Fourier coefficients obtained $a_n$ (also called FDs) are normalized by using the DC component of the transform as

$$f = [\frac{|FD_2|}{|FD_1|}, \frac{|FD_3|}{|FD_1|}, ... \frac{|FD_{N-1}|}{|FD_1|}] \qquad (16)$$

Where FD1 is the DC component, and $FD_N$ are the Fourier coefficients.

FDs are invariant to translation since the shape descriptors used to derive FDs are invariant to translation. Scale invariance is achieved through normalization of the Fourier coefficients and rotation invariance is achieved by considering only the magnitude values of FDs and ignoring the phase information. The FDs are widely used for shapes recognitions in various fields [40, 42].

*Hausdorff Distance (HD)*

The Hausdorff distance [43] performs shape matching on a point-to-point basis. It is mainly used for template matching and measuring the similarity between shapes. For two shapes represented as two sets of points A = {a1, a2, … $a_p$} and B = {b1, b2, b3,…$b_q$}, the HD between these two shapes A and B is given by

$$H(A,B) = \max\{h(A,B), h(B,A)\} \tag{17}$$

Where

$$h(A,B) = \max_{a \in A} \min_{b \in B} \|a - b\| \tag{18}$$

It has the advantage of being able to match shapes partially. However it is not invariant to translation, scale and rotation. So matching shapes involve overlaying them on each other at different positions, sizes and orientations which make the matching process very time consuming and expensive. It is also very sensitive to noise and slight variations. Efforts have been made to improve this scheme for shape matching. One such effort is the shape context.

*Shape Context*

Shape recognition using shape contexts [44] is an enhancement to the classic Hausdorff distance based methods. It captures a global feature, called shape context (SC) for each corresponding point. The context features of all corresponding points are then matched for matching shapes with each other. To obtain the SC at a boundary point p, the distances of p to all the other boundary points are calculated. The length r and direction θ of the distance vectors are then quantized to form a histogram map which represents the point p. The histogram of each point is compressed and combined together to form the context of the shape. In order to make the histogram sensitive to nearby points than to those of farther points, these distance vectors are placed into log-polar space. The process of shape context computation has been illustrated in the Figure 8 given below.

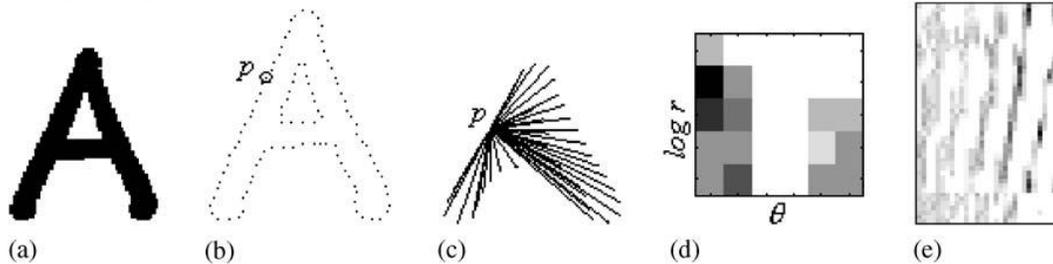

**Figure 8.** Shape context. (a) Shape of a Character; (b) edge map of (a); (c) shape boundary point (a) and all the vectors originating from p; (d) the log-polar histogram of the vectors in (c), the histogram is the context of point p; (e) the context map of shape (a), each row of the context map is the flattened histogram of each point context, each row represents as boundary point. (re-printed from [44])

The matching process involves matching of the two shape contexts which is a matrix based matching.

**Structural Shape Descriptors**

For structural shape description, the shape's contour is broken down into basic components called primitives. These techniques vary in the choice of primitives and their organization for shape representation. Some of the basic and well known structural shape representation schemes are presented in this section.

*Chain Code (CC)*

It was first introduced by Freeman [45]. It is a method of representing shape's contour as a string of positive integers representing the direction of change in the boundary as it is traversed.

445                                     JOURNAL OF PLATFORM TECHNOLOGY VOL. 2, NO. 4, DECEMBER 2014A grid based approximation of the shape contour is obtained and then a 4-directional or 8-directional chain code can be generated by traversing the approximated shape boundary. If the CC is to be used for recognition it should be independent of the starting point and also invariant to rotation. To achieve so a first difference of the CC is computed by taking the difference of the CC value from the previous value modulo n. (n is the connectivity i.e. 4 or 8). This code is known as the *differential CC* and is invariant to rotation provided that the rotation angle is a multiple of 90 degrees. To achieve starting point invariance this differential CC is rotated such that code yields the lowest integer value lexicographically. The resulting code is called *shape number*. The chain code is very much sensitive to noise and hence it is not for higher level analysis like polygonal approximation.

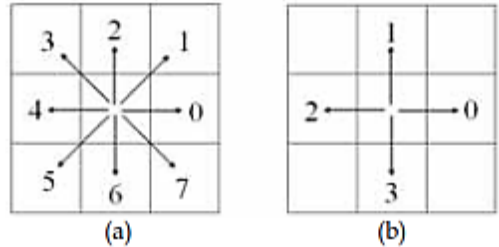

**Figure 9.** Illustrations of Basic Chain Codes (a) 8-directional CC (8-connectivity); (b) 4-directional CC (4-connectivity)

*Smooth Curve Decomposition*
The smoothed shape boundary is broken down into curve segments to obtain the primitives. These primitives are called tokens. The features of these tokens is its maximum curvature and orientation. In the given Figure 10, the first value is the maximum curvature and the second value is orientation.

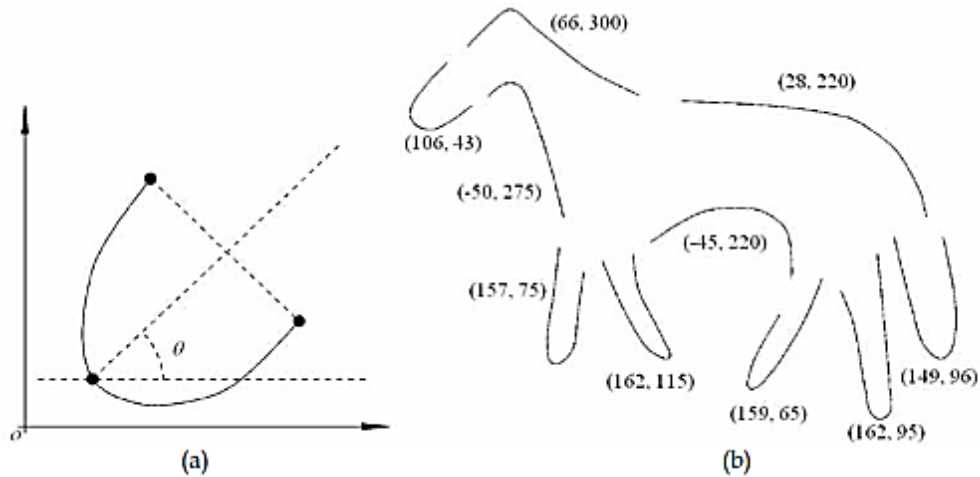

**Figure 10.** Illustration of Smooth curve decomposition (a) θ is the token's orientation; (b) Example

For shape matching based on this scheme, a weighted Euclidean distance is used. It has been proved that shape matching based on smooth curve decomposition representation is robust to partially occulted shapes and also invariant to translation, scale and rotation [46].

*Graph Based Shape Representation*
The use of graph for shape representation has the advantage that it is flexible and tolerant to scaling, rotation and translation. A graph based shape description and matching scheme is being presented as follows.

In [47], a method has been presented which uses basic geometric properties extracted from binary shapes and represents these properties as an Attributed Relational Graph (ARG). In order to achieve rotation and scale invariance, the extracted attributes are relatively computed and then normalized. The feature extraction and graph generation process is as follows.

The raster binary image representing the shape is converted to vector image by using polygonal approximation on the shape contour. The attributes of the primitives' i.e. relative length and angle are stored in a chained list. For thin or linear shapes, quadrilaterals are formed using vectors that are close to each other having similar slope. In this case thinning is not used because it introduces rough branches at crossings and line junctions. Primitive attributes like relative length of the vectors are represented as graph nodes and the relationship between nodes i.e. relative angles form the edges of the graph.

Hence a binary shape is represented as a 4-tuple:

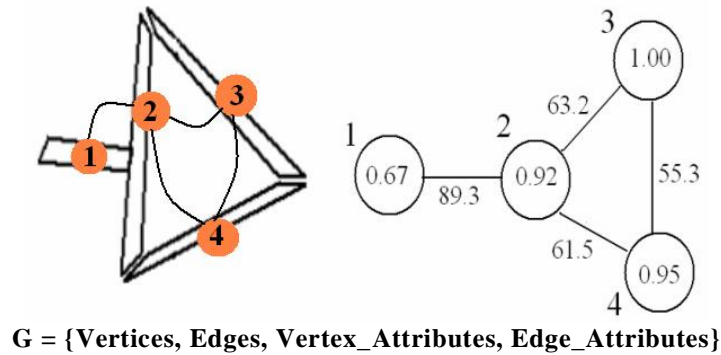

**G = {Vertices, Edges, Vertex_Attributes, Edge_Attributes}**

**Figure 11.** A graph derived from the shape

Two shapes represented as ARGs are compared using a greedy algorithm which looks for optimal mapping between the two graphs followed by computation of a similarity score.

## 3. Conclusion

An overview of image content description methods has been presented. The low-level features for visual content derived from color, texture and shape has been highlighted. The aim of this study was to present fundamental characteristics for representing images in compact, effective and discriminative manner that will allow their efficient indexing and retrieval. The most notable color descriptors along with their strengths and weaknesses have been presented. Some of the dominant texture description and shape description methods were also highlighted. For the purpose of image indexing and retrieval, these characteristics are widely used. Current research focuses on improving these descriptors, specifying new ways to describe visual content and to use powerful discriminative features along with state-of-the-art machine learning algorithms to reduce the so called semantic gap.

## 4. Acknowledgment

This research is supported by (1) the ICT R&D program of MSIP/IITP. [2014(R0112-14-1014), The Development of Open Platform for Service of Convergence Contents], and (2) The Basic Science Research Program through the National Research Foundation of Korea (NRF) funded by the Ministry of Education (2013R1A1A2012904).